\documentclass{iauc}
\usepackage{graphicx}

\title[Galaxy Morphological loop] 
{Are violent events responsible of a Galaxy Morphological loop?}

\author[Scannapieco et al.]   
{Cecilia Scannapieco$^1$
, A.V. Smith Castelli$^{1,2}$ 
\and P.B. Tissera$^1$}

\affiliation{$^1$Instituto de Astronom\'{\i}a y F\'{\i}sica del Espacio, Conicet - UBA, 
\break CC 67 - Suc 28 (1428) Ciudad de Buenos Aires, Argentina \break
email: cecilia@iafe.uba.ar, patricia@iafe.uba.ar\\[\affilskip]
$^2$Facultad de Ciencias Astron\'omicas y Geof\'{\i}sicas, Universidad Nacional de La Plata, \break Paseo del Bosque s/n (1900) La Plata,
Argentina \break
email: asmith@fcaglp.unlp.edu.ar}

\pubyear{2005}
\volume{198} 
\date{?? and in revised form ??}
\jname{Near-Field Cosmology with Dwarf Elliptical Galaxies}
\editors{H. Jerjen and B. Binggeli, eds.}
\begin{document}

\maketitle

\begin{abstract}
We use cosmological SPH simulations to  investigate the effects of mergers and 
interactions on the formation of the bulge and disc components of galactic
systems.
We find that secular evolution during mergers seems to be a key process in the     
formation of stable disc-bulge systems with observational counterparts
and contributes to establish the fundamental relations
observed in galaxies. 
Our findings suggest that the secular evolution phase couples the formation
mechanisms of the bulge and disc components.
According to our results, depending on the particular
stability properties and merger parameters, violents events could drive a morphological
loop in which the outcome could be a disc or a spheroid.

\end{abstract}

\firstsection 
\section{Introduction}

The understanding of galaxy formation remains a primary goal
in astrophysics. 
Within the current accepted scenario of structure formation, small
galaxies are built up first and larger objects are assembled through the
merging of less massive building blocks, in a hierarchical way.
In this context, mergers and interactions play a crucial role
modifying the properties of galaxies such as their mass distribution 
and star formation activity  and 
affecting the subsequent evolution of the systems.

The formation of disc-like systems is still a matter of debate
and several scenarios have been proposed in order to explain the
growth  of these two components and  their interactions.
The simplest scenarios currently accepted for the formation of the
bulges are two: the first one establishes that the bulge and the
disc are formed independently (Andredakis et al. 1995), and  the 
second one assumes that the disc forms first and the bulge
emerges from it as a consequence of gas inflow during a period of secular
evolution (Courteau et al. 1996).

Numerical simulations are a powerful tool to tackle galaxy formation since they
can account for the joint evolution of baryons and dark matter
in a cosmological framework.
We use Smooth Particle Hydrodynamical (SPH) simulations to investigate
the effects of mergers and interactions  on the determination of the structural
properties of galaxies. 

\section{Analysis and Discussion}

We analize galactic objects identified from numerical SPH simulations
that take into account the gravitational
and hydrodynamical evolution of the matter in Cold Dark Matter universes.
We run two simulations with $N=64^3$ (S1) and  $N=2\times 80^3$ (S2) total number
of particles and the following cosmological parameters:
$\Lambda=0$, $h=0.5$, $\Omega_b=0.10$ (S1)and $\Lambda=0.7$,  $h=0.7$, $\Omega_b=0.04$ (S2), where 
($H=100 h {\rm \ km \ s^{-1} \  Mpc^{-1}}$). The simulated boxes
represent cubic volumes of $5h^{-1}$ (S1) and $10h^{-1}$  (S2) Mpc length.

We select galaxy-like objects (GLOs) at $z=0$ by using a density-contrast
criterium and follow their evolution with look-back time,
constructing their star formation  and merger histories (see Scannapieco \& Tissera 2003 for details).
With the aim at investigating the effects of mergers on the structural
properties of the systems we perform
bulge-disc decompositions to the projected mass surface density of the GLOs at $z=0$, 
and of the progenitors during merger events, assuming an exponential
profile for the discs and the S\'ersic law for the bulges.
Galaxy-like objects identified in our simulations are found to
have structural parameters similar to those of observed spiral galaxies
(S\'aiz et al. 2001).


Tissera et al. (2002) showed that 
during merger events, early gas inflows can be triggered depending
on the stability properties of the systems.
Those with shallower potential wells are more likely to suffer gas
inflows since they can be more strongly  perturbed
from the incoming satellite.
Secular evolution phases feed the bulge components which
significantly contribute to the stability of the discs.
The actual fusion of the baryonic cores increases the star formation
activity which in turn affect the subsequent properties of the systems.

Understanding the effects of these two processes (i.e. secular
evolution and fusion) on the growth of
the disc and bulge components in galaxies can help us to 
enlighten the picture of galaxy formation. To this end we analize
the evolution of the mass distribution of galaxy-like objects during merger events
by studying the variation of their structural parameters.
We find that secular evolution during the orbital decay phase produces 
the largest changes in the structural parameters of the
systems: bulges get larger with respect to the discs 
and with more exponential profiles. These changes are found to be 
more important in unstable systems with
shallower potential wells. Fusions  introduce further modifications in the opposite 
direction although they are less important. 

We also find that secular evolution phases contribute to establish
the observed  fundamental relations: after secular evolution, bulges tend to satisfy the observed 
Fundamental Plane and the disc components the Tully-Fisher relation. Fusions
may add some dispersion but do not significantly change them
(Smith Castelli et al. 2005).
Departures from the Fundamental Relations could be indicating a lack of secular phases.

Our findings suggest that interactions play a fundamental role
being able to transform
discs into spheroids. Discs can also survive violent events,
being either partially destroyed or just feeded by the colliding
object. 
The stability properties
of the initial system seem to be a fundamental point in the response
of the disc to the incoming satellite.
Other factors such as orbital parameters and  relative masses of the
colliding objects would also affect the determination of the structural
properties of the outcome systems.
The different possible histories of evolution of galactic systems in hierarchical
universes could potentially explain the diversity of galaxies in our Universe.

\end{document}